\documentclass[12pt]{article}

\usepackage{epsfig}

\newcommand{\beq}{\begin{equation}}
\newcommand{\eeq}{\end{equation}}

\newcommand{\ve}[1]{{\vec{\bf #1}}}

\newcommand{\dive}{\mbox{$\nabla \cdot$}}

\begin{document}

\title{
Exact nonlinear analytic Vlasov-Maxwell tangential equilibria with arbitrary density and temperature profiles}

\author{F. Mottez, \\
Centre d'\'etude des Environnements Terrestre et Plan\'etaires (CETP),  \\
10-12 Av. de l'Europe,
78140 V\'elizy, France. \\
(e-mail: fabrice.mottez@cetp.ipsl.fr)} 


\maketitle

\begin{abstract} 
The tangential layers are characterized by a bulk plasma velocity 
and a magnetic field that are perpendicular to the gradient direction.
They have been extensively described in the frame of the Magneto-Hydro-Dynamic (MHD) theory.
But the MHD theory does not look inside the transition region
if the transition has a size of a few ion gyroradii. 
A series of kinetic tangential 
equilibria, valid for a collisionless plasma is presented. 
These equilibria are exact analytical solutions of the Maxwell-Vlasov equations.
The particle distribution functions
are sums of an infinite number of elementary
functions parametrized by a vector potential.
Examples of equilibria relevant to space plasmas are shown. 
A model for the deep and sharp density depletions observed in 
the auroral zone of the Earth is proposed. Tangential equilibria are also relevant for 
the study of planetary environments and of
remote astrophysical plasmas.
\end{abstract} 

{\tiny
\begin{copyright}
Copyright (2003) American Institute of Physics. This article may be downloaded 
for personal use only. Any other use requires prior permission of 
the author and the American Institute of Physics. 
The following article appeared in Phys. Plasmas, Vol 10, 7, pp2501,2508, and may be found at 
http://link.aip.org/link/?php/10/2501.
\end{copyright}
}

\newpage

\section*{I. INTRODUCTION}

The swedish Viking spacecraft orbiting in the auroral region 
of the magnetosphere discovered deep plasma density depletions 
\cite{Higers et al 1992a}. Their size along the 
magnetic field direction reach thousands of kilometers.
They can be considered as channels parallel to the magnetic field.
These observation were confirmed by measurements onboard the 
Fast Auroral SnapshoT explorer (FAST) spacecraft
\cite{Strangeway et al 1998}.
The data showed a lot of plasma activity 
in and around these regions, particularly plasma acceleration 
(at the origin of the polar auroras) \cite{McFadden et al 1998}, 
electromagnetic radiation \cite{Higers 1992b}
and a high level of electrostatic turbulence.
Most of the studies focused on the acceleration 
\cite{Genot et al 1999}, \cite{Genot et al 2000}
and the nature of the turbulence \cite{Roux et al 1993}, \cite{Pottelette et al 2001}
\cite{Genot et al 2001}.
But surprisingly, little work has been devoted to the structure 
of these plasma cavities. Some theoretical works have been devoted to their origin,
but none has tackled the question : are auroral plasma cavities in equilibrium,
or do they vanish a soon as the cause of their formation disapears ?

The aim of the work presented in this paper is to construct
a kinetic model of the auroral plasma cavities. 
But as a very general solution to this problem is found, the
field of applications appears much wider.
I  present here a family of equilibrium solutions that can be used in many domains
of the physics of plasmas. They will be named as tangential equilibria because,
as in the MHD tangential discontinuities, the magnetic field and the bulk
velocities are tangential to the plane $(y,z)$ of invariance.

Previous works on tangential equilibria \cite{Harris 1962}, \cite{Channell 1976}
have shown the existence of simple isothermal
solutions that can be computed analytically. Other works, mainly focused on the 
study of the Earth magnetopause \cite{Roth et al 1996}, have shown
solutions that satisfy a larger class of constraints, but they cannot be constructed
analytically.

Most of these equilibria a based, for each species, on distribution functions that 
are products of an exponential function of the total energy and of $p_z^2$, by
an arbitrary $g$ function of $p_y$,
\begin{equation} \label{densite-distri-g-general0}
f= (\frac{\alpha_{z} \alpha_{\perp }^2}{\pi^3})^{1/2} 
[\exp{(-\alpha_{z} v_{z}^2  -\alpha_{\perp } v_{\perp }^2)} ] \,
g(mv_y +q A_y(x)),
\end{equation}
where $\alpha_{z}=m/2T_{z}=1/v_{tz}^2$, 
$\alpha_{\perp }=m/2T_{\perp }=1/v_{t \perp }^2$,
are the reciprocals of the squared thermal velocities ; 
$v_{z}=p_{z}/m$
and $v_{\perp}^2=v_{x}^2+v_{y}^2$. 
If $g(\infty) \sim 1$,
then $f$ is a bi-maxellian distribution for $x \sim \infty$.

Krall and Rosenbluth, \cite{Krall Rosenbluth 1963} have derived a solution of
the linearized kinetic equations, by setting $g(p_y)$ as a linear function. 
They used this simple solution for the study of gradient instabilities.

Many authors (reviewed by Roth {\em et al.} \cite{Roth et al 1996}), 
who studied tangential discontinuities,  considered 
solutions  with $g$ functions that involve 
error functions.
Most of these studies where developed for the study of 
the equilibrium of the Earth magnetopause.
They include an electric field $E_x$ and no charge neutrality.
There is no analytical solution. The differential equations on $E_x$ and $A$ are solved 
through a numerical integration.

Channell \cite{Channell 1976} considered solutions 
with no electric field. He showed that 
\beq \label{densite-g-Jy-dn-dAy}
 J_y= T_{\perp} \frac{dn}{dA_y}.  
\eeq
A more general formula will be shown in Sec. II.

In the same paper, Channell studied a few examples
of solutions with simple but nonlinear $g$ functions. In particular, he 
showed that there exists a solution with $g(p_y)=\exp{(-\eta p_y^2)}$, 
where $\eta$ is a constant number.

Harris \cite{Harris 1962} built an equilibrium model of a current sheet. 
His model was widely used later for the study of the tearing mode instability.
The Harris current sheet equilibrium corresponds to the case 
$g(p_y)=\exp{(\nu p_y)}$.

All the solutions mentionned above are based on elementary $g$ functions. 
They do not let the possibility of setting {\em a priori} a density profile, as
in the more simple MHD and bi-fluid models.
We can build a distribution function that is the sum of such elementary 
distribution functions and assert that each elementary part of the distribution
corresponds to a family of trapped particles. 

Because of the nonlinearity of the equations, the density and temperature profiles
associated to a sum of families of trapped particles are not simple combinations
of the profiles associated to single elementary particle distributions.
Nethertheless, using a numerical equation solver,
Roth {\em et al}. \cite{Roth et al 1996} added up to four elementary distributions functions
(all of them were built with error functions). Differential equations were solved numerically.

I show in this paper how to combine an infinite number of elementary 
solutions and build a plasma that is made of a continuum  
of famillies of trapped particles.
The plasma equilibria
presented in this paper are computed analytically. They do not have the high degree 
of generality of those shown in Roth {\em et al.} \cite{Roth et al 1996}, 
(because I impose a null electric field), but they
are more general than those of Harris \cite{Harris 1962}, and Channell \cite{Channell 1976}. In particular, the plasma
does not have to be isothermal.

The equations of the kinetic model are developed 
and solved in Sec. II.
The properties of the solutions and some particular cases
are briefly analysed in Sec. III. 
An equilibrium model for the deep non isothermal density depletions of 
the Earth auroral zone is presented.
The last section is a discussion about 
other applications and further developments.

\section*{II. FORMULATION OF THE MAXWELL-VLASOV MODEL} 
We consider a monodimensional equilibrium : $\partial_t =0$ and $
\nabla =(d_x,0,0)$. We choose the $z$ direction along the constant 
direction of the magnetic field
$\ve{B}=(0,0,B_z(x))$, ($\nabla \cdot B =0$ is therefore ensured).
The magnetic field derives from a vector potential  
$\ve{A}(x)= (0,A_y(x),0)$ such that $B_z=d_x A_y$. This vector potential
satisfies the Lorentz gauge (that is $\dive \ve{A}=0$ in the stationary case).
The $x$ direction is called the normal direction, 
the $y,z$ plane is the tangential plane. 

In a monodimensional equilibrium, the
particles are confined between two points where $v_x=0$. In the case of a null electric field, 
the size $\Delta x$ of this area is 
\beq
\Delta x = 2 v_{\perp} /<\omega_c>
\eeq
where
$<\omega_c>=(q/m \Delta x) \int_{x_1}^{x_2} B_z(x)$.
Absolutely no restriction on the variations of $B_x$ is made. 

As we look for a monodimensional  equilibrium, the
total energy $E=\frac{mv^2}{2}+q\Phi$ of any particle and the generalized
momentum $p_y=mv_y + q A_y$ and $p_z=mv_z$ are the invariants of the motion.
Any distribution function of the form $f=f(E,p_y,p_z)$
is a solution of the Valsov equation. All the kinetic solutions studied in this
paper will be of that form.
The other character imposed on the solutions is a null electric field, 
and therefore no charge separation.

The basic idea is to decompose linearly the $g$ function 
(already mentionned in Eq. \ref{densite-distri-g-general0} ) over a set of elementary 
distribution functions that correspond to an analytical equilibrium solution. 
As will be shown in this section, if the sum is parametrized by a parameter $a$
that correspond to shifts of the elementary distribution functions in the
space of the potential vectors (and not in the space of the configurations),
there is an analytical solution.

For instance, we can show that a continuous linear combination of Channell and 
Harris-like $g$ distribution functions given by 
\beq \label{distribution-gy0}
g({p_y})= n_0 + \int_{a_{1}}^{a_{2}} da \; n_g(a) 
\exp{[-\eta(a) (\frac{p_y}{m}-\frac{q}{m}a)^2 +\nu(a)(\frac{p_y}{m}-\frac{q}{m}a)]},
\eeq
leads to an equilibrium that can be computed analytically.

The functions $n_g(a)$, $\eta(a)>0$, and $\nu(a)$ are defined almost arbitrarily 
and $n_0$ is a constant scalar.
Each of these three functions provide a degree of freedom and it is possible, 
as in  MHD and bi-fluid models, to control the plasma density profile. 
But the distribution of the particles depends on the energy 
through an exponential factor that does not give any freedom to set the temperature profile.
Therefore, instead of using \ref{densite-distri-g-general0} and \ref{distribution-gy0},
let us set : 
\beq \label{distribution-general}
f= \int_{a_1}^{a_2} da \; (\frac{\alpha_{z}(a) \alpha_{\perp }^2(a)}{\pi})^{1/2} 
e^{(-\alpha_{z}(a) (v_{z}-u_{z})^2  -\alpha_{\perp }(a) v_{\perp }^2)} 
(n_0 + g_{a}({p_y}))
\eeq
and 
\beq \label{distribution-gy}
g_{a}({p_y})= n_g(a) 
e^{-\eta(a) (\frac{p_y}{m}-\frac{q}{m}a)^2 +\nu(a)(\frac{p_y}{m}-\frac{q}{m}a)}.
\eeq
With these distribution functions, a control of the plasma density profile (through 
$n_g(a)$) and 
of the temperature profile of each species 
(through $\alpha_{z}(a)$ and $\alpha_{\perp }(a)$) is possible.
The parameter $a$ is homogeneous to a vector potential. 

In order not to overload the equations, we do not express systematically 
the dependance of the parameters on $a$ and $x$ hereafter.

To solve the charge neutrality equation ($n_i=n_e$) 
and the Amp\`ere equation
\beq \label{densite-ampere}
- \frac{d^2}{d x^2} A_y=\mu_0 J_y, 
\eeq
we need to compute the contribution of each species $s$ to the charge 
density,
\begin{equation} \label{densite-densite-cinetique-g}
n_s =  \int_{a_1}^{a_2} da \;  (\frac{\alpha_{\perp}}{\pi})^{1/2} 
\int  dv_y \; \exp{( -\alpha_{\perp} v_y^2)} 
g(mv_y +q A_y(x)).
\end{equation}
and to the current density,
\begin{eqnarray} \label{densite-courant-cinetique-g}
J_{ys} &=& \int_{a_1}^{a_2} da \; q  (\frac{\alpha_{\perp}}{\pi})^{1/2}
\int dv_y  \; v_y \exp{( -\alpha_{\perp} v_y^2)} 
g(mv_y +q A_y(x)) \\
&=& \int_{a_1}^{a_2} da \; q  (\frac{1}{ 4 \pi \alpha_{\perp}})^{1/2}
\int dv_y  \;  \exp{( -\alpha_{\perp} v_y^2)} 
g'(mv_y +q A_y(x)) 
\end{eqnarray}
where $g'$ is the derivative of $g$. 
These charge and current densities can be computed explicitely. 
We reorder Eq. (\ref{distribution-general}) and (\ref{distribution-gy})
to separate the terms 
that depend on $v_y$ and $v_y^2$,
\begin{eqnarray}
\label{densite-s3-f-ordered}
f= & \int_{a_1}^{a_2}da \;  \sqrt{\frac{\alpha_{\perp}^2 \alpha_z}{\pi^3}} 
 e^{-\alpha_{\perp} v_x^2 -\alpha_z (v_z-u_z)^2} \times \\ 
\nonumber
&   
 \{n_0 e^{-\alpha_{\perp} v_y^2}  +    
n_g(a) 
e^{-\eta(a) [\frac{q}{m}(A_y(x) -a)]^2 + \nu(a) \frac{q}{m} (A_y(x)-a)}
e^{-P(a) v_y^2 +2 Q(a) v_y}
 \},
\end{eqnarray}
with 
\beq
P(a) = \alpha_{\perp} + \eta(a) \;\;\; \mbox{et}\;\; 
Q(a)=\frac{1}{2} \nu(a) -\eta(a) \frac{q}{m} (A_y(x)-a)
\eeq
The contribution of each species $s$ to the charge and to the current densities
are 
\beq
n_s=n_0 +\int_{a_1}^{a_2} da \; n_g \sqrt{\frac{\alpha_{\perp}}{\pi}} e^{-\eta (q/m)^2 A_y^2} S_{0,P,Q}
\eeq
where $S_{0,P,Q}$ is defined by 
\beq
S_{0,P,Q}= \int_{-\infty}^{+\infty} dv_y e^{(-P v_y^2 +2 Q v_y)} 
=
\sqrt{\frac{\pi}{P}} e^{(\frac{Q^2}{P})} \\
\eeq
and 
\beq
J_{ys}=\int_{a_1}^{a_2} da \; q  n_g \sqrt{\frac{\alpha_{\perp}}{\pi}} e^{-\eta (q/m)^2 A_y^2} 
S_{1,P,Q},
\eeq
where 
\beq
S_{1,P,Q} =\int_{-\infty}^{+\infty} dv_y \, v_y e^{(-P v_y^2 +2 Q v_y)}=
\frac{Q}{P} \sqrt{\frac{\pi}{P}} \exp{(\frac{Q^2}{P})}.
\eeq
Hence,
\begin{eqnarray} \label{densite-s3-densite-charge}
n_s&=&n_0 + \int_{a_1}^{a_2} da \,
n_{g} \sqrt{\frac{\alpha_{\perp}}{\alpha_{\perp}+\eta}} 
e^{\frac{\nu^2}{4(\alpha_{\perp}+ \eta)}}
e^{- \frac{\alpha_{\perp}}{\alpha_{\perp}+\eta} 
[\frac{q}{m} (A_y-a)][\eta \frac{q}{m} (A_y-a) -\nu]} \\ \nonumber
J_{ys}&=&-q  \int_{a_1}^{a_2} da \,
\frac{n_g}{2} \sqrt{\frac{\alpha_{\perp}}{\alpha_{\perp}+\eta}} 
e^{\frac{\nu^2}{4(\alpha_{\perp}+ \eta)}}
\frac{ 1 }{\alpha_{\perp}+\eta} \times \\
\label{densite-s3-densite-courant}
 &&[-2 \eta \frac{q}{m} (A_y-a)+\nu] 
 e^{- \frac{\alpha_{\perp}}{\alpha_{\perp}+\eta} 
[\frac{q}{m} (A_y-a)][\eta \frac{q}{m} (A_y-a) -\nu]}.
\end{eqnarray}
Let  
\begin{eqnarray} \label{densite-s3-xi}
\xi(a) &=& \frac{\alpha_{\perp}(a) \eta(a)}{\alpha_{\perp}(a)+\eta(a)}(\frac{q}{m})^2   \\
\label{densite-s3-delta}
\delta(a)&=&\frac{\alpha_{\perp}(a) \nu(a)}{\alpha_{\perp}(a)+\eta(a)}(\frac{q}{m})        \\
\label{densite-s3-N0}
N_0(a)&=&  n_g(a) \sqrt{\frac{\alpha_{\perp}(a)}{\alpha_{\perp}(a)+\eta(a)}}
e^{\frac{\nu^2}{4(\alpha_{\perp}+ \eta)}}.
\end{eqnarray}
The contribution of each particle species to the particle density is:
\beq \label{densite-s3-densite-charge-reduit}
n_s(x)=n_0 + \int_{a_1}^{a_2} da \, N_0
e^{- \xi (A_y-a)^2 +\delta (A_y-a)}
\eeq
Let us set 
\beq
n_a(x)=N_0(a) e^{- \xi(a) (A_y(x)-a)^2 +\delta(a) (A_y(x)-a)},
\eeq
then 
\beq \label{densite-s3-densite-charge-resume}
n_s(x)=n_0 + \int_{a_1}^{a_2} da \, n_a(x)
\eeq
If the following equations are satisfied for each value of $a$, 
\begin{eqnarray}
\xi(a)_{ions} = \xi(a)_{electrons} \\
\delta(a)_{ions} = \delta(a)_{electrons} \\
N_0(a)_{ions}= N_0(a)_{electrons},
\end{eqnarray}
then, the charge neutrality is satisfied.
It is easy to show that 
one can freely choose the functions $\alpha_{\perp}, \, \eta, \nu, \, n_g$ 
that correspond to a particle species, then compute the $\xi$, $\delta$ and $N_0$ functions
and find the $\eta, \, \nu, \, n_g$ functions of the other species
that are associated with the same $\xi$, $\delta$ and $N_0$. 
There is only one restriction: $\eta$ must be positive.
If we choose electron parameters, the condition 
for a positive ion $\eta_i$ parameter is, for each value of $a$
\beq \label{densite-g-nonlineaire-extreme-eta-0}
\frac{m_e}{\eta_e(a)} > 2 [(\frac{m_i}{m_e})T_{\perp i}(a) +T_{\perp e}(a)]
\eeq
which requires small enough values of $\eta_e(a)$. 
Noticing that $\eta$ homogeneous to the reciprocal of a squared velocity $\eta=v_{\eta}^2$, 
the above relation writes 
\beq \label{densite-g-nonlineaire-extreme-v-eta-0}
v_{\eta}^2 > (\frac{m_i}{m_e} \frac{T_i}{T_e}+1) v_{te}^2
\eeq
Let us call $\eta_c(a)= v_c(a)^{-2}$ the value that corresponds to an equality in 
Eq. (\ref{densite-g-nonlineaire-extreme-eta-0}) and (\ref{densite-g-nonlineaire-extreme-v-eta-0}).
The case of $\eta_e=\eta_c$ corresponds to an infinite value of $\eta_i$. 
We shall see in the applications that because of the condition $\eta < \eta_c$, $v_{\eta}$ can reach very high values, 
and it is necessary to remember that $v_{\eta}$ is not an actual velocity, 
it does not correspond to a propagation phenomenon. 

Let us define  
\beq \label{densite-s3-k(a)}
k(a)=  -\mu_0 N_0(a) 
(\frac{m_e}{\alpha_{\perp e}(a)}+\frac{m_i}{\alpha_{\perp i}(a)})
= -2 \mu_0 N_0(a)(T_{\perp e}(a)+T_{\perp i}(a)).
\eeq
Negative values of $k(a)$ correspond to a positive value of $n(a)$, and therefore
imply an increase of the plasma density. On the contrary, positive values
of $k(a)$ induce a reduction of the density.
The total current density is the sum of the ion and the electron currents,
\beq \label{densite-s3-densite-courant-reduit}
J_y(x)=-\frac{1}{\mu_0} \int_{a_1}^{a_2} da \; \frac{k}{2} 
[-2 \xi (A_y -a) + \delta] e^{-\xi (A_y(x)-a)^2 +\delta (A_y(x)-a)}.
\eeq
Let us notice that Eq. (\ref{densite-s3-densite-charge-reduit}),
(\ref{densite-s3-k(a)}), and (\ref{densite-s3-densite-courant-reduit}) imply 
\beq
J_y= \int_{a_1}^{a_2} da \; T_{\perp}(a) \frac{d n_a}{d A_y}.
\eeq
The above relation is a generalization of Eq. (\ref{densite-g-Jy-dn-dAy}) to non isothermal equilibria.\\
The Amp\`ere equation becomes
\beq \label{densite-s3-equation-courant-ordre2}
\frac{d^2 A_y(x)}{dx^2}=
\int_{a_1}^{a_2} da \; \frac{k(a)}{2} 
[-2 \xi (A_y(x) -a) + \delta] e^{-\xi (A_y(x)-a)^2 +\delta (A_y(x)-a)}.
\eeq
We can deduce a first order equation 
\beq \label{densite-s3-equation-courant-ordre1}
B_z(x)^2= (\frac{d A_y}{dx})^2 =C +  
\int_{a_1}^{a_2} da \,
k(a) e^{-\xi(a) (A_y-a)^2+\delta (A_y(x)-a)}.
\eeq
The vector potential is the solution of
\beq \label{densite-s3-solution-x-Ay}
x = s \int_{A_y(0)}^{A_y(x)} 
\frac{dA}{\sqrt{C+  \int_{a_1}^{a_2} da \,
k(a) \exp{[-\xi(a) (A-a)^2+\delta(a) (A-a)]}}}.
\eeq
The sign of the magnetic field is given by $s=\pm 1$, and 
$C$ is an integration constant. The parameters 
$k$, $\xi$, and $\delta$ are arbitrary functions of $a$.
The solution has a physical meaning if it is defined for any value of $x$:
the integral in Eq. (\ref{densite-s3-solution-x-Ay}) 
considered as a function of $A_y$
must be able to vary from $-\infty$ to $+\infty$.

Equation (\ref{densite-s3-equation-courant-ordre1}) 
can be modified in order to provide
an integro-differential equation in $B_z(x)$. Let $y$ defined by 
$A_y(y)=a$ be the new integration variable.
Then, 
\beq \label{densite-s3-magnetique-ordre1}
B_z(x)^2 = C + \int_{y_1}^{y_2} dy B_z(y) k(y) 
e^{ 
-\xi(y) [\int_{y}^{x}B_z(u) du]^2+\delta(y) [\int_{y}^{x}B_z(u) du] 
 }.
\eeq
The same process applied on Eq. (\ref{densite-s3-densite-charge-reduit})
provides
\beq \label{densite-s3-densite-ordre1}
n(x) = n_0 + \int_{y_1}^{y_2} dy B_z(y) N_0(y) 
e^{ 
-\xi(y) [\int_{y}^{x}B_z(u) du]^2+\delta(y) [\int_{y}^{x}B_z(u) du] 
 }.
\eeq
These equations will be used in section III.B to analyse the asymptotic behaviour of the solutions.

\section*{III. EXAMPLES AND PROPERTIES OF TANGENTIAL EQUILIBRIA} \label{section-properties}

\subsection*{A. Examples of elementary solutions with parameters relevant to space plasmas}
The trivial case of a uniform plasma can be easily recovered with several sets of parameters 
(including $n_g \ne 0$ and $\xi \ne 0$).
The case of the Harris current sheet \cite{Harris 1962}, 
reviewed in details in \cite{Roth et al 1996}, can be recovered with
$\xi=0$, $\nu \ne 0$ and $n(a)=n_g \delta_0$, where $\delta_0$ is the Dirac distribution.

Here are a few examples of equilibria where  $\nu=0$, $n_g=n_c \delta_0$, 
and where the temperature functions are constant.
They correspond to the case $g(p_y) = n_c \exp{(- \eta (p_y/m)^2)}$ already mentionned
by Channell \cite{Channell 1976}. The figure given in Channell's paper
to illustrate this case 
displays the equilibrium of an evanescent plasma.
We briefly show in this section 
a few examples of equilibria of finite size current layers
in non evanscent plasmas,
in order to emphasize a few interesting properties not discussed in Channell's paper.

Figure 1 shows an example of a structure that is computed with parameters
typical of high altitude auroral plasmas : asymptotic magnetic field
$\sqrt{C}=3000$ nT, $n_0=10 \mbox{cm}^{-3}$, $T_i=T_e=10$ eV, 
and $\eta_e=1.53 \, 10^{-16} (m/s)^{-2}=0.99 \eta_c$, very close to the 
critical value imposed by
Eq. (\ref{densite-g-nonlineaire-extreme-eta-0}).
A high negative value $n_c=-8 \mbox{cm}^{-3}$ allows for a deep density depletion.
The abcsissa is normalized to the ion Larmor radius which is
$\rho_L = 152$ m. We can see on Fig. 1 that the size of the structure is about 
$4 \rho_L \sim 600$ m. Such a narrow and deep structure 
(the density in the cavity is only 20\% of the density of the surrounding plasma)
can be used in a first approximation for the density depletions encountered in the
high altitude auroral zone. 

The solution displayed in Fig. 2 corresponds to the limiting case for $\eta$, obtained for  
the weakly magnetized interplanetary plasma encountered in the solar wind at one 
astronomical unit : 
$\sqrt{C}=5$ nT, $n_0=10 \mbox{cm}^{-3}$, $T_i=T_e=100$ eV, 
and $\eta_e=1.53 \, 10^{-17} (m/s)^{-2}=0.99 \eta_c$. The ion Larmor radius 
is $280$ km. We can notice that for the same density 
depletion, the variation of the magnetic field is much higher than in the case 
of a higly magnetized plasma. We also notice that the total size of the structure
hardly exceed one ion Larmor radius.

As the size is of the order of the ion Larmor radius,
such solutions cannot be described through the fluid theories.
Of course, we can also build very large density structures, if we choose a small
value of ${\eta}$ in order to get a small value of $\xi$. Such large 
structures can more simply be described by fluid models.
Figure 3 shows the solution with $\eta =10^{-20} (m/s)^{-2}\sim 10^{-3} \eta_c $. 
The structure is much larger 
but has the same amplitude. Comparing Fig. 2 and Fig. 3, 
we can notice that, contrary to the case of solitons, 
there is no correlation between the size and the amplitude of the structures.

\subsection*{B. Asymptotic behaviour of the solutions} 

We can caracterize two kinds of non trivial elementary solutions : 
those with $\eta \ne 0$, and those where $\eta = 0$ (and therefore $\nu \ne 0$, otherwise the solution is trivial).
Those of the first kind correspond to localized solutions, they are centered around an abscissa $x$ such as
$A_y(x) \sim a$. Therefore, when the distribution is the 
infinite sum of elementary functions indexed by $a$ as in Eq. (\ref{densite-s3-densite-charge-resume}), 
the parameter $a$ indicates where the 
density $n_a(x)$ is non negligible. 
Let us consider an equilibria where, for $x \sim \infty$, the magnetic field has a finite value. 
Then, the potential vector goes to infinity. Taking $a_1 \sim -\infty$ and $a_2 \sim +\infty$ means that 
the plasma distribution will be influenced by some of those elementary 
functions even when $x \sim \infty$. On the contrary, equilibria defined with finite values 
of $a_1$ and $a_2$ (and $B(\infty) \ne 0$) means that 
for $x \sim \infty$, the equilibria will converge toward the trivial solution of a uniform plasma with the density $n_0$ 
that appears in Eq. (\ref{distribution-general}).

The case of a null magnetic field at $x \sim \infty$ (less useful for space plasmas applications) 
is different. Let us consider the case of  $B(+\infty)=0$. For $x \sim +\infty$, 
the vector potential has a constant value. 
Let us call $A_{\infty}$ this value. Then, as $A$ cannot exceed $A_{\infty}$, 
the main integral in Eq. (\ref{densite-s3-solution-x-Ay})
is divergent for $A_y(x)=A_{\infty}$ (otherwise the equilibria would not be defined for any value of $x$). 
As the elementary solutions parametrized 
by $a$ are non negligible only for abscissas $x$ such that $A_y(x)$ is close to $a$, it is not 
necessary to consider values or $a_2$ that exceed notably $A_{\infty}$ : they would have no influence on 
the value of the integral in Eq. (\ref{densite-s3-solution-x-Ay}). In brief, for equilibria with
$B(\infty)=0$, the range of values where the parameter $a$ is significant is bounded, 
$a_1$ and $a_2$ do not have to be 
infinite.

In the Harris model, the magnetic field reverses and has two opposite finite values for 
$x \sim - \infty$ and $x \sim + \infty$. We shall see
that there exist non elementary solutions where the magnetic field has finite values
on both sides, and that these values have not necessarily the same absolute value.

We consider the case $a_1=-\infty$ and $a_2=+\infty$.
Let $u^-=\lim_{x \rightarrow -\infty} u$ be the limit of 
any physical value $u$ when $x$ tends to $-\infty$,
and $u^+=\lim_{x \rightarrow +\infty} u$.
The integral in Eq. (\ref{densite-s3-magnetique-ordre1}), where $y_1=-\infty$
and $y_2=+\infty$ can be cut in three parts : an integral $I1$ from
$y_1$ to $x-\Lambda$, an integral $I2$ from $x-\Lambda$ to $x+\Lambda$,
and an integral $I3$ from $x+\Lambda$ to $y_2$.
We will consider large values of $x$ (positive or negative), and large values
of $\Lambda$.

For $\Lambda$ large enough,  $y$ has large negative values, 
\begin{eqnarray}
I1 
&\sim& 
\int_{y_1}^{x-\Lambda} dy B_z^- k^- 
\exp{(-\xi^- (B_z^-)^2 (x-y)^2-\delta^- B_z^-(x-y)  )} \\
&\sim& 
-\frac{k^-}{\delta^-} [\exp{\delta^- B^- \Lambda}-\exp{\delta^- B^- y_1}]
\end{eqnarray}
This function tends to zero for $y_1 \rightarrow -\infty$ if $\delta^- B^-<0$,
and diverges if $\delta^- B^->0$. In the case $\delta^- =0$, 
\beq
I1 \sim
\int_{y_1}^{x-\Lambda} dy B_z^- k^- 
\exp{(-\xi^- (B_z^-)^2 (x-y)^2 )} 
\eeq
tends toward zero.
Similarily, $\lim_{y_1 \rightarrow -\infty} I3 = 0$ if $\delta^+ B^+ \ge 0$.
With the same kind of technique, we can show that, for large negative values of $x$,
\begin{eqnarray}
I2 
&\sim& 
k^- \sqrt{\frac{\pi}{\xi^-}} \sigma_{B_z^-} \exp{\frac{(\delta^-)^2}{4\xi^-}},
\end{eqnarray}
where $\sigma_{B_z^-}$ is the sign of $B_z^-$.
Provided that $\delta^+ B^+ \ge 0$, and $\delta^- B^- \le 0$,
the asymptotic expansion of Eq. (\ref{densite-s3-magnetique-ordre1}) shows that
\beq
(B_z^-)^2 = 
C +
k^- \sqrt{\frac{\pi}{\xi^-}} \sigma_{(B_z^-)} \exp{\frac{(\delta^-)^2}{4\xi^-}}.
\eeq
Computing an asymptotic value of $I2$ for $x \rightarrow +\infty$
bring a similar  relation for $(B_z^+)$.
The same method used with Eq. (\ref{densite-s3-densite-ordre1})  provides
the asymptotic value
of the  contribution of each species to the particle density:
\beq
n^{+}=  n_0 + N_0^{+} \sigma_{(B_z^+)} 
          \sqrt{\frac{\pi}{\xi^+}} \exp{\frac{(\delta^+)^2}{4\xi^+}}.
\eeq
The equation for $n^{-}$ is analogous.
As $d_x n (\infty)=0$, we deduce
$J_y^+=J_y^-=0$.
There is no current density carried by the plasma for $x \sim \infty$.
This result can also be found with an asymptotic expansion 
of Eq. (\ref{densite-s3-densite-courant-reduit}).
Moreover, Eq. (\ref{densite-s3-densite-courant-reduit}) shows that the 
ion and  electron contribution to
the current density have the same sign, their mean velocities are therefore 
opposite. As there is no current for $x \sim \infty$, for each species 
$v^+=v^-=0$. The plasma does not flow perpendicularly to the magnetic field 
at $x \sim \infty$.

\subsection*{C. Examples of solutions with and without field reversal}

Figure 4 shows an example of equilibrium where the magnetic field amplitude 
and the density are not the same for 
$x \sim - \infty$ and $x \sim + \infty$. We have set 
constant values $\eta=1.53 \;  10^{-17}(m/s)^{-2}$ close to $\eta_c$, $\nu=0$, and a function 
$n_g(a)=640 \mbox{atanh} (1000 a)$. The temperature 
is 10 eV. With $C=(5nT)^2$, the asymptotic magnetic field amplitude 
is 5nT. The ion Larmor radius is $\rho_i \sim 0.2 \, 10^{6}$m.
The variable $a$ is homogeneous to $B_z$ times $x$, therefore $1000 a \sim x/\rho_i$ 
is of the order
of magnitude of the absissa $x$ divided by the ion Larmor radius.
The factor $640$ that multiplies the inverse hyberbolic tangent 
function was set heuristically in order to
get a variation of density of the order of 60\%.
The relatively low value of $\sqrt{C}=5$ nT 
allows for high relative variations of the magnetic field amplitude. 
The asymptotic values of the magnetic field and of the density 
fit the analytical formulas given in section III.B.

Figure 5  shows an example where $\eta = 0$. 
We have chosen  a function 
$n_g(a)=4200 e^{-(1000 a)^2}$. The temperature 
is 10 eV, and $\sqrt{C}=5$ nT.
The value $\nu=10^{-10}$ was chosen, because, in the exponential, the 
factor $\nu a \sim \nu B_z x$ is again of the order of $x/\rho_i$.
We can see on Fig. 5 that the sign of the magnetic field changes
for $x \sim - \infty$ and $x \sim + \infty$. The reversal occur 
for a value of $A_y$ that cancels the square root in Eq. (\ref{densite-s3-solution-x-Ay}).
Although the magnetic field is reversed on both sides, 
this is not a Harris equilibrium. 

\subsection*{D. Example of a non isothermal plasma cavity in the 
Earth auroral zone} 

Are the cavities built in the Earth auroral zone in an equilibrium state of the plasma ?

Hilgers {\em et al.} \cite{Higers et al 1992a} made a careful analysis of the Langmuir probe
current measurements onboard the swedish Viking satellite in low plasma densities.
They showed a case, taken at an altitude of 7000 km, of a deep  auroral cavity.
The density reach less than one particle
per cubic centimeter, that is less than 10 \% of the surrounding plasma
density. The electron plasma temperature in the cavity (although not precisely measured) 
is of the order of 1 keV, 
compared to 1 eV outside the cavity. The boundaries of the cavity are sharp, 
Hilgers {\em et al}. measured 1.4 km, of the order of a few ion Larmor radii. 
The magnetic field amplitude is  $B_{z0}=6800$ nT. 

Apart from the temperature gradient, all these features have been qualitatively reproduced 
in the elementary equilibria shown in Fig. 1. 
In order to reproduce a cavity with sharp temperature gradients, 
an elementary solution cannot be used. 
Let $\rho_i$ be the Larmor radius outside the cavity, it corresponds to the cold palsma (1 eV), 
its value is 21 m. Inside the cavity, the plasma temperature is 1 keV, and the ion Larmor radius is
$\rho_{i,hot}=660$m$=32 \rho_i$. 
We choose an electron temperature function 
$T_e(a)=1+1000 \exp{(-(a/B_{z0} 32 \rho_i)^2)}$ that varies 
from 1 to 1000 eV on a scale $\Delta x \sim 32 \rho_i \sim \rho_{i,hot}$.
The density is controled through the constant scalar $n_0=10$ cm$^{-3}$, 
and the $n_g$ function, 
$n_g(a)=-n_{g0} \exp{(-(a/B_{z0} 32 \rho_i)^2)}$.
The scalar $n_{g0}$ is set (heuristically) to 50000 in order to have 
one particle per cubic centimeter inside the cavity. The $\nu$ function is null.
The $\eta$ function is chosen is order to get a sharp cavity, 
that is with a value close to the $\eta_c$ function:  
$\eta(a)=0,98 \eta_c (a)$ have values of the order of $10^{-15} (m/s)^{-2}$. Actually, the 
equilibria is not very sensitive on the value of $\eta$. Taking a lower value like $\eta(a)=0.7 \eta_c (a)$ 
(and a slightly lower value of $n_{g0}$) bring similar results. 

The result, displayed on Fig. 6, is in quantitative agreement with the prescriptions given 
by the observations. The size of the gradient is $70 \rho_i =1400$m, as measured by Hilgers {\em et al}.

We can notice on figure 6 that the magnetic field amplitude variation $\Delta B$ in the cavity is about 0.5 nT. 
A similar equilibria (not displayed)
set with a lower value of the ambient magnetic field ($B_{z0}=300$ nT) bring a larger variation: $\Delta B=5$ nT. 

We conclude from this short study that deep plasma cavities can be equilibrium structures of the 
Earth auroral plasma. Therefore, they won't be destroyed immediately after the
extinction of their cause. It is however interesting to know 
if this equilibrium is stable. The theoretical treatement 
of this question goes beyhond the scope of this paper. 
I will only give a few hints
in favour of the stability of the auroral cavities. 

The Auroral zone of the Earth is a radio source. 
The power of the emissions, called Auroral Kilometric Radiation (AKR),
can reach 10 MW. 
The Viking spacecraft has gone through the sources of AKR: 
the waves are emitted inside the cavities, where they take their
free energy from the hot rarefied plasma \cite{Roux et al 1993}.
The waves are strongly refracted at the edges of the AKR source,
and it is probable that a part of the AKR is guided inside the cavities
\cite{deFeraudy 1987}.
Auroral kilometric radiations can be observed on time scales of 10 minutes.
The cavities that contain AKR sources are expected to last at least for the same duration.
Therefore, two situations are possible : 
(1) the cavities are the consequence of a phenomena that lasts 
for tens of minutes, or 
(2) they are generated by more transient phenomena and they are stable.

An another hint about the stability of hot plasma cavities 
comes from numerical simulations carried by Genot {\em et al}. 
\cite{Genot et al 2001}. 
The simulations start with an isothermal (a few eV) auroral cavity. The cavity is stable. 
An Alfv\'en wave is added. The interaction of the Alfv\'en wave with the cavity
triggers strong electron acceleration and turbulence. When the free energy is completely 
removed from the Alfv\'en wave by the accelerated electrons, the plasma inside
the cavity is hot, and the surrounding plasma is cold.
We are in the situation shown in Fig 6. The simulations show that the heated cavity,
which is still very deep and sharp, is stable.

Most of the models about the generation of auroral cavities invoke strong kinetic Alfv\'en 
Waves (SKAW) \cite{Wu et al 1997} \cite{Shukla et al 1999}, observed onboard the Freja 
and FAST satellites  \cite{Louarn et al 1994},
\cite{Chaston et al 1999}. 
The associated magnetic field perturbation $\Delta B$ 
is of the order of 50 nT. But the observations made onboard 
FAST show that the SKAW's are mainly observed in the cusp and the polar cap boundary layer,
while the cavities are observed in the lower latitude auroral region where 
the magnetic fluctuations
$\Delta B$ do not exceed 5 nT. It is possible that the cavities are built in the 
polar cap boundary layer by SKAWs, they subsist after the
disapearance of the Alfv\'enic perturbations and finally they 
are gently convected to lower latitudes
by the large scale convection electric field. This suppose of course the stability of the
auroral cavities.  An other scenario is that the cavities are created 
through Field Line Resonance (FLR) \cite{Lotko et al 1998} 
directly in the auroral region. The 
$\Delta B \sim $ 5 nT  (on a time scale that corresponds to the crossing
of a cavity by a spacecraft) associated to the FLR is compatible with 
the value shown in Fig. 6. The equilibria shown in Fig. 6 may be 
a relevant short scale Vlasov-Maxwell description
of the part of the FLR that is in the high altitude auroral zone 
(the FLR has been modeled on a global scale
in the multifluid approximation up to now).

\section*{IV. CONCLUSION AND FURTHER DEVLOPPMENTS}

This paper presents a large class of monodimensional analytic kinetic equilibria.
They are based on particle distribution functions Eq. (\ref{distribution-general}) 
that depend on a set of almost arbitrary functions. These distribution
functions are solutions of the Vlasov equation, 
and the vector potential can be computed through the evaluation of the integral
function given in Eq. (\ref{densite-s3-solution-x-Ay}).
In some particular cases, like the uniform plasma and the Harris current sheet,
this integral is a combination of elementary mathematical functions.
In the other cases, its numerical evaluation is straightforward.

Unlike most of the kinetic 
tangential equilibria given in the litterature,
the freedom in the choice of the density profile, altough more difficult to control,
is almost as large as with bi-fluid models. The kinetic 
equilibria have the great advantage of giving a complete description of the distribution
function (not provided with the fluid theory), 
and can describe equilibria where strong gradients
develop on the scale of a few ion Larmor radii.
These solutions can be used as initial conditions in particle in cell
and Vlasov numerical simulations.

Most of the authors who build models and are cited in this paper   
have based their elementary solutions
on distribution functions that depend on the invariants of the motion. 
Some of them combined a small number of such elementary
solutions and asserted that each 
of these elementary solutions
correspond to a family of trapped particles. The distribution function 
given in Eq. (\ref{distribution-general}) is the superposition, 
not of a finite number, but of
a continuum of families of trapped particles. 
This is why we have a very large degree of freedom
in the choice of the density and temperature profiles.
For the first time, we show that such a supperposition of families 
of trapped particles lead to an analytically integrable equation, 
whose solution is given in Eq. (\ref{densite-s3-solution-x-Ay})

The equilibria discussed in section III.B where the plasma
is uniform for $x \sim \infty$ can be compared to the solutions of 
the jump equations 
developed in the frame of the MHD theory.
The equilibria developed in the present paper are characterized 
by jumps of the density and of the
magnetic field, the normal velocity is null as well as the normal magnetic field.
They therefore belong to the family of tangential discontinuities.
But they concern only a sub-category: 
the magnetic field direction is uniform and
the solutions have no velocity shear: the velocities
are equal to zero at $x \sim -\infty$ and $x \sim +\infty$, 
even if they can take other values
at finite distances.
These restrictions are not required for general tangential discontinuities.

Building a kinetic model, inspired from the present results,
of a tangential discontinuity where the magnetic field can turn
is straighforward and will be presented in a forthcomming paper in order to 
analyse experimental data provided by the Cluster satellites.

A kinetic model of a tangential equilibria whith a velocity shift 
($x \sim -\infty$ and $x \sim +\infty$) is not compatible with exact 
charge neutrality. A quasi neutral model would do. It could be computed, as 
a first order perturbation in $n_i-n_e$ added to 
the charge neutral equilibria presented
in this paper.

The examples given in the present paper come from the magnetospheric and solar wind 
physics because
these media offers the opportunity of in situ observations of 
non collisional astrophysical plasmas.
Tangential equilibria can be used to describe  the magnetopause 
(present around all the magnetized planets with an atmosphere),
the distant neutral sheet, the Earth auroral cavities, and some 
density fluctuations of the 
solar wind \cite{Safrankova et al 2000}. 

But tangential equilibria do not exist only in the terrestrial environnment.
They exist whenever plasmas with different origins and velocities meet. 
Such situations exist in many astrophysical plasmas.
Theoretical models (based on MHD and plasma multifluid theories) of the
boundary of the Heliosphere show that it is constituted of a termination shock, 
followed by the heliopause, that is a tangential discontinuity, an hydrogen wall,
and possibly an heliospheric shock \cite{Zank 1999}. 
Tangential discontinuities also exist around other stars. 
M\"uller {\em et al.} \cite{Muller et al 2001} studied the interaction of 
the very active 
binary star $\lambda$ Andromedae, and of the neaby star $\epsilon$ Indi 
with the interstellar medium.
They found, as in the case of the heliopause, the existence of four boundaries, 
one of them,
the asteropause being a tangential discontinuity.
It is clear that tangential layers exist in other astrophysical plasmas where 
they may play a very important role, as frontiers, or in 
acceleration, heating and radiative processes. 

The studies of the magnetospheric plasmas
have shown that most of the heating and acceleration phenomena 
can be only explained through 
kinetic processes.
So far, most of the remote astrophysical plasmas have been studied through 
MHD or multifluid theories.  The understanding of acceleration processes in remote 
astrophysical plasma might as well require kinetic models. The equilibria 
presented in this paper might be a good start for some of those studies.

%
\newpage

\section*{ACKNOWLEDGMENTS}
The author gratefully acknowledges stimulating and usefull discussions with 
G\'erard Belmont and Alain Roux.

%

\newpage

\begin{figure}[htp]
\centering
\vspace{1cm}
\psfig{file=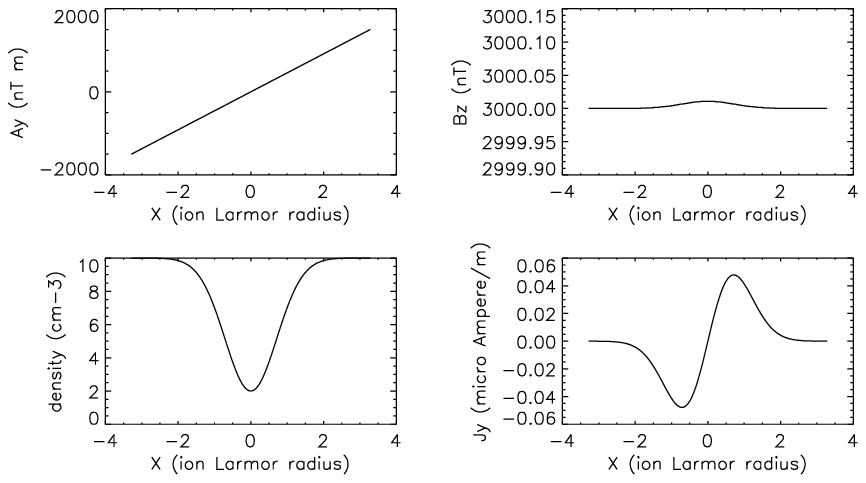,width=12cm}
\caption{ FIG. 1. A deep and narrow density depletion in a highly magnetized plasma.
The parameters of this equilibrium are given in Sec. III.A.}
\end{figure}

\begin{figure}[htp]
\centering
\vspace{1cm}
\psfig{file=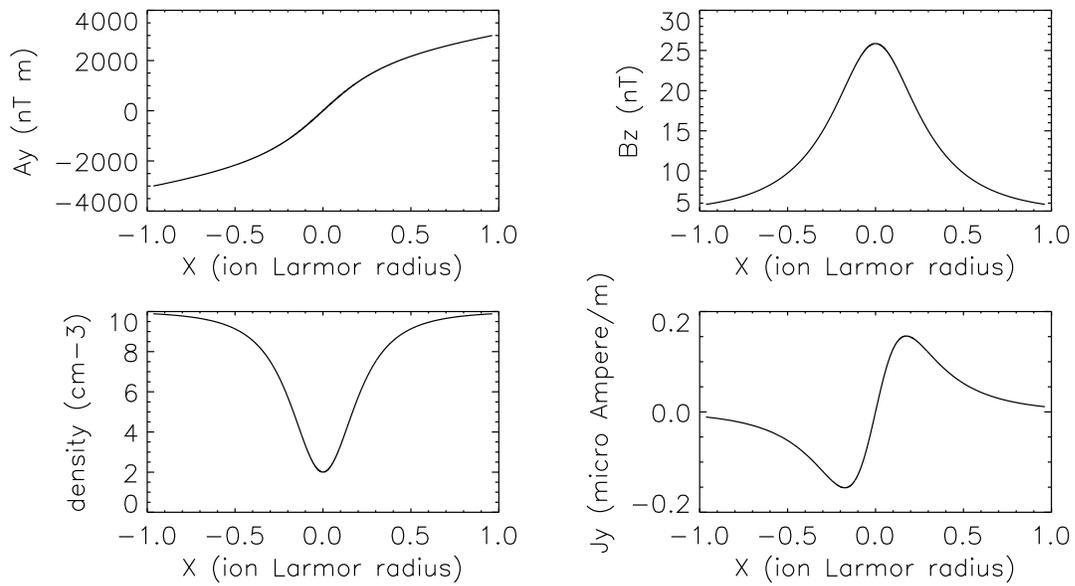,width=12cm}
\caption{FIG. 2. A deep and narrow density depletion plasma with a weak magnetic field.
The parameters are the same as for Fig. 1, exepct for $C$ that is much weaker.
}
\end{figure}

\begin{figure}[htbp]
\centering
\vspace{1cm}
\psfig{file=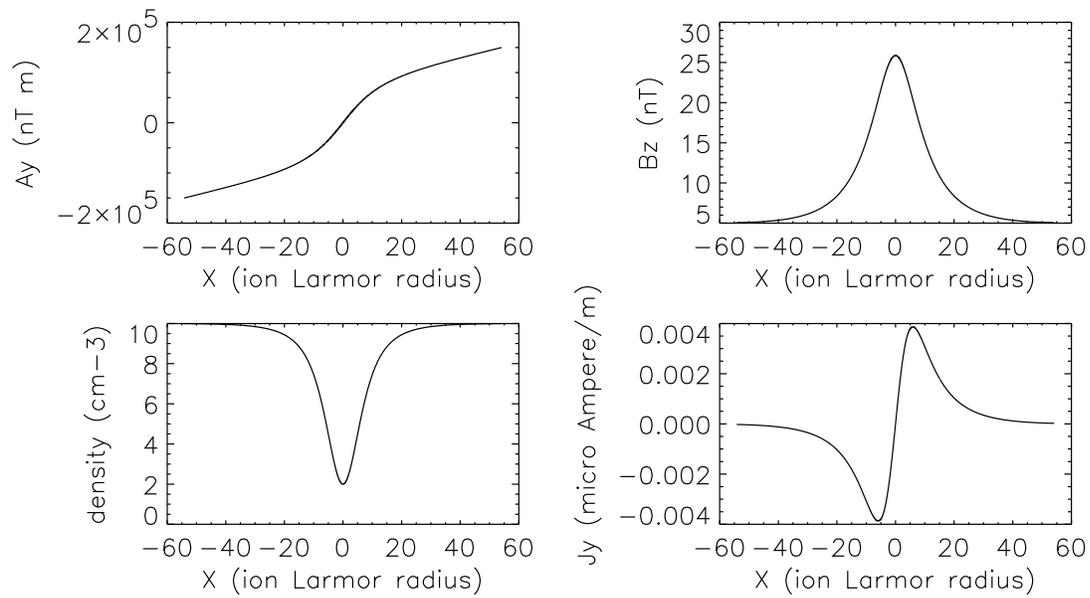,width=12cm}
\caption{FIG. 3. A deep and large density depletion, the only difference with Fig. 2 is
a larger value of $\eta$.
See Sec. III.A..}
\end{figure}

\begin{figure}[htp]
\centering
\vspace{1cm}
\psfig{file=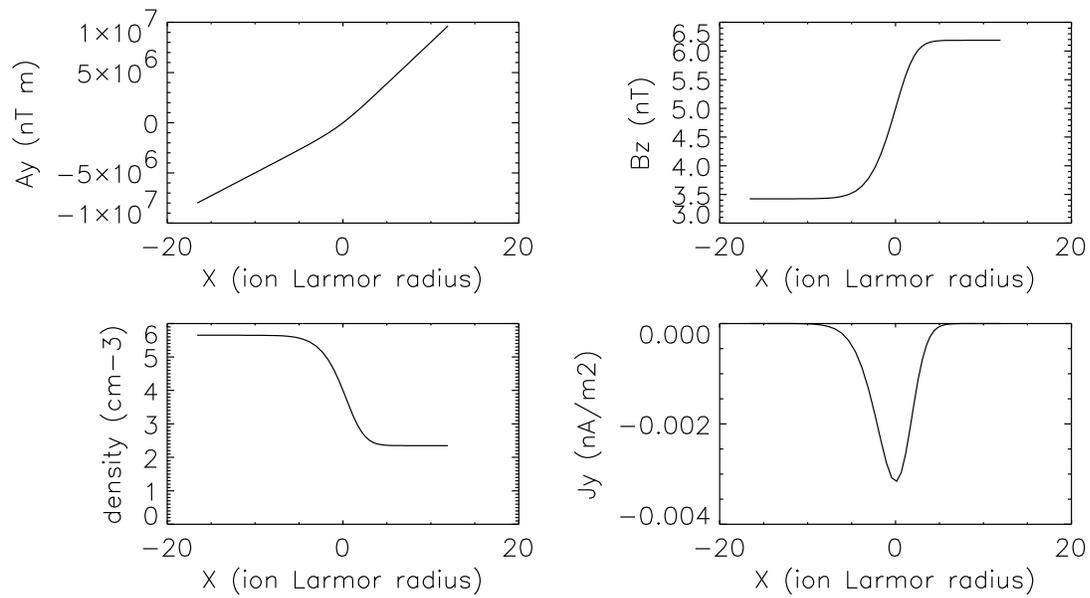,width=12cm}
\caption{FIG. 4. An example of non symetric equilibrium where $n_g$ is a function of $a$.
See Sec.  III.C for details.}
\end{figure}

\begin{figure}[htp]
\centering
\vspace{1cm}
\psfig{file=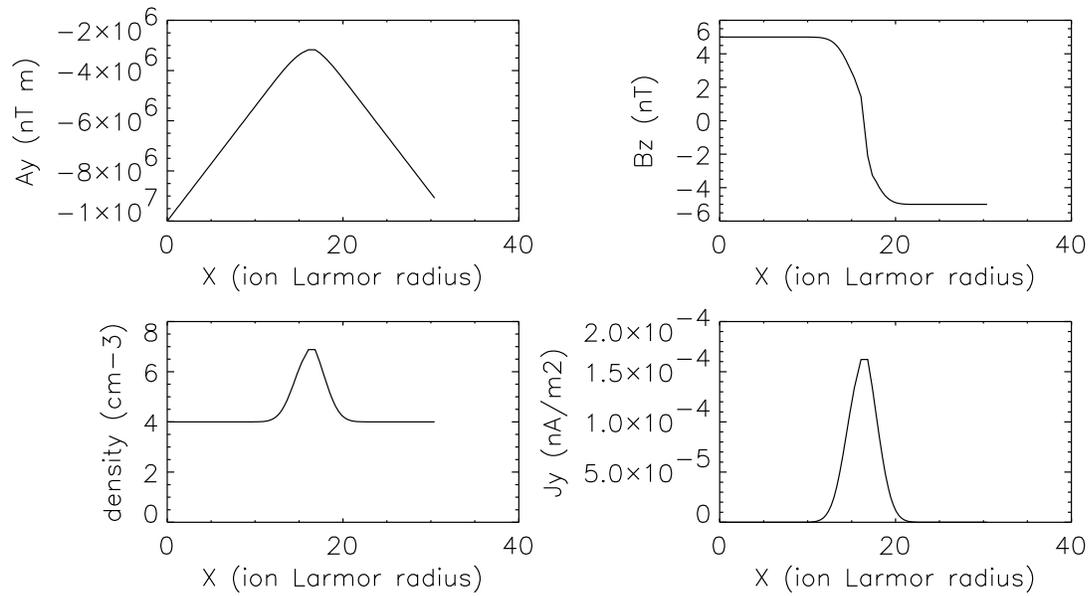,width=12cm}
\caption{FIG. 5. An exemple of an equilibrium with $\eta=0$ and 
a reversal of the magnetic field. Details are given in Sec III.C.}
\end{figure}

\begin{figure}[htp]
\centering
\vspace{1cm}
\psfig{file=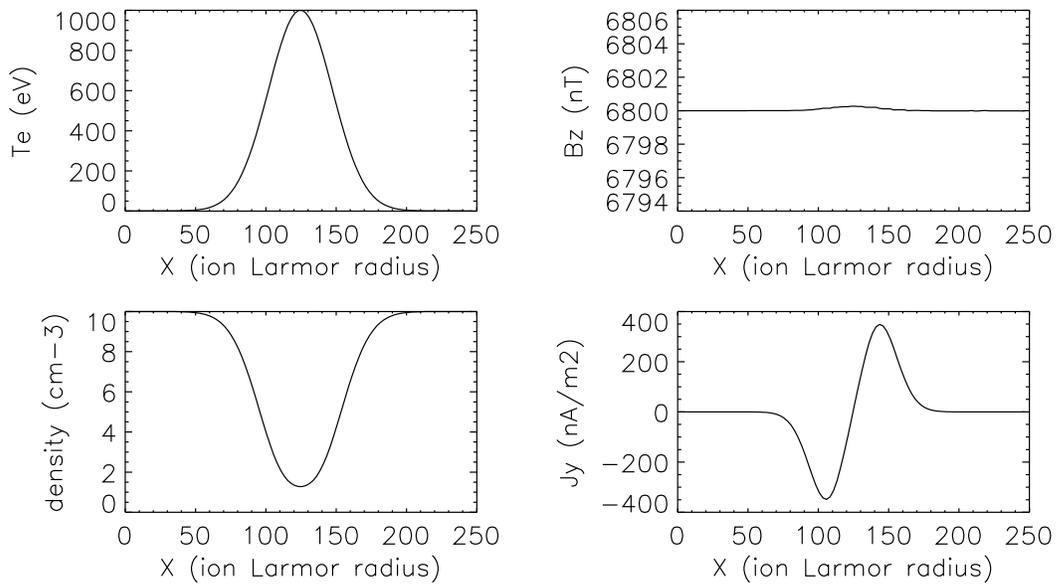,width=12cm}
\caption{FIG. 6. An example of a deep plasma cavity containing hot electrons (1 keV) 
surrounded by a cold (1 eV) highly magnetized plasma.  
See  section III.D for details.}
\end{figure}

\end{document}